\documentclass[floatfix,reprint,amsmath,amssymb,aps,prl,superscriptaddress,sort]{revtex4-2}

\usepackage{graphicx}
\usepackage{bm}
\usepackage{float}
\usepackage{physics}
\usepackage[colorlinks=true, allcolors=blue]{hyperref}
\usepackage[export]{adjustbox}

\usepackage[caption=false]{subfig}
\usepackage{graphicx}
\usepackage{cleveref}

\newcommand{\expvalue}[1]{\langle #1 \rangle}

\begin{document}

\title{Majorana physics in a Luttinger liquid with attractive interactions}

\author{Francesco Debortoli}
\email{f.debortoli@lmu.de}
\affiliation{Department of Physics and Arnold Sommerfeld Center for Theoretical Physics (ASC), Ludwig-Maximilians-Universität München, Theresienstr. 37, München D-80333, Germany}
\affiliation{Munich Center for Quantum Science and Technology (MCQST), Schellingstr. 4, München D-80799, Germany}
\affiliation{Max-Planck-Institute for Quantum Optics, Hans-Kopfermann-Str. 1, Garching D-85748, Germany}
\author{Nitya Cuzzuol}
\affiliation{Institute for Condensed Matter Physics and Complex Systems, DISAT, Politecnico di Torino, Torino I-10129, Italy}
\author{Luca Barbiero}
\affiliation{Institute for Condensed Matter Physics and Complex Systems, DISAT, Politecnico di Torino, Torino I-10129, Italy}
\author{Fabian Grusdt}
\affiliation{Department of Physics and Arnold Sommerfeld Center for Theoretical Physics (ASC), Ludwig-Maximilians-Universität München, Theresienstr. 37, München D-80333, Germany}
\affiliation{Munich Center for Quantum Science and Technology (MCQST), Schellingstr. 4, München D-80799, Germany}

\date{\today}

\begin{abstract}
    Majorana zero modes are the hallmark of topological superconductivity. In one-dimensional systems, these zero modes are usually introduced in the context of gapped, mean-field models that do not conserve particle number, such as the Kitaev chain. By non-locally encoding a conventional fermion across spatially separated Majorana zero modes, these systems become inherently immune to local decoherence. In this work, we show that signatures of Majorana edge physics persist in a number-conserving, gapless Luttinger liquid of spinless fermions with short-range attractive interactions. We identify the two-point correlator as a sharp diagnostic, revealing an edge-to-edge revival whose sign depends on the fermion-number parity. This revival is robust in the thermodynamic limit, and persists in the excited states of the system and at different fillings. A simple particle-hole ansatz for the ground state of the system with an odd number of fermions captures the physics of the system for a wide range of interaction strengths, interpolating between the free-fermion limit and the strongly interacting Majorana regime. Finally, we propose a concrete protocol to realize this model with ultracold dipolar molecules or atoms in an optical lattice, and to detect the revival via beam-splitter interferometry, opening an experimental route to Majorana physics beyond the conventional gapped-superconductor paradigm.
\end{abstract}

\maketitle

\begin{figure}[ht]
    \centering
    \captionsetup[subfigure]{labelformat=empty}
    \begin{minipage}[c]{0.49\columnwidth}
        \centering
        \subfloat{\hspace{0.4em}\includegraphics[width=\linewidth, valign=c]{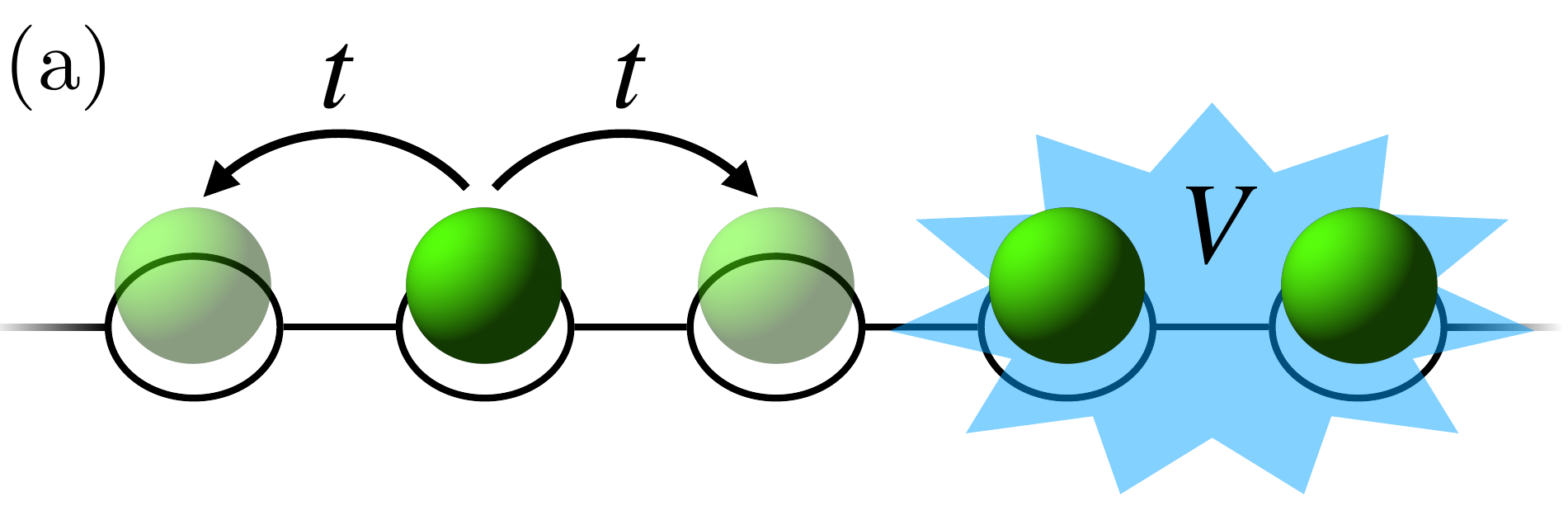}\label{fig:processes}}
        
        \vspace{0.8em} % Adjust vertical spacing between the two left images
        
        \subfloat{\includegraphics[width=\linewidth, valign=c]{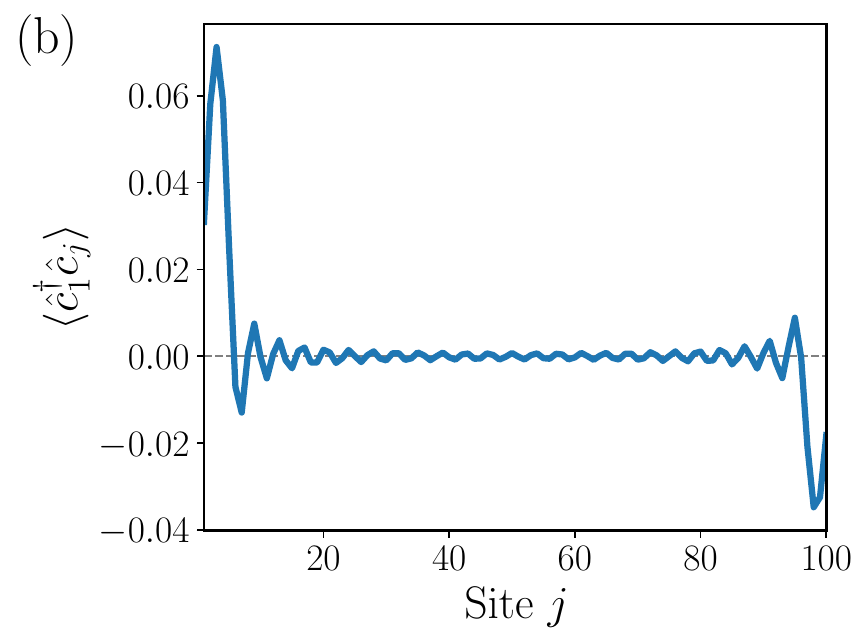}\label{fig:revival}}
    \end{minipage}
    \hfill
    \begin{minipage}[c]{0.49\columnwidth}
        \centering
        \subfloat{\hspace{0.6em}\includegraphics[width=\linewidth, valign=c]{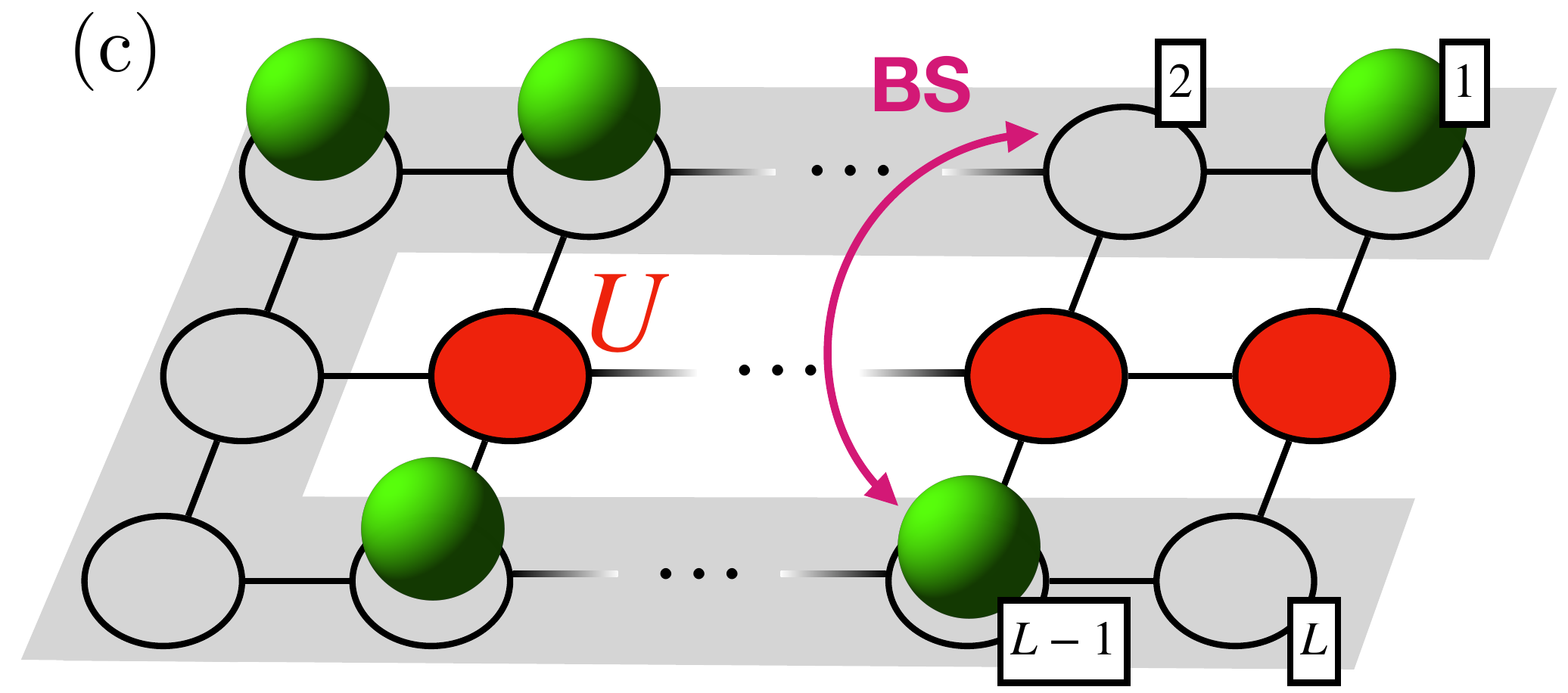}\label{fig:horseshoe_setup}}
        
        \vspace{-1.2em} % Adjust vertical spacing between the two right images
        
        \subfloat{\includegraphics[width=\linewidth, valign=c]{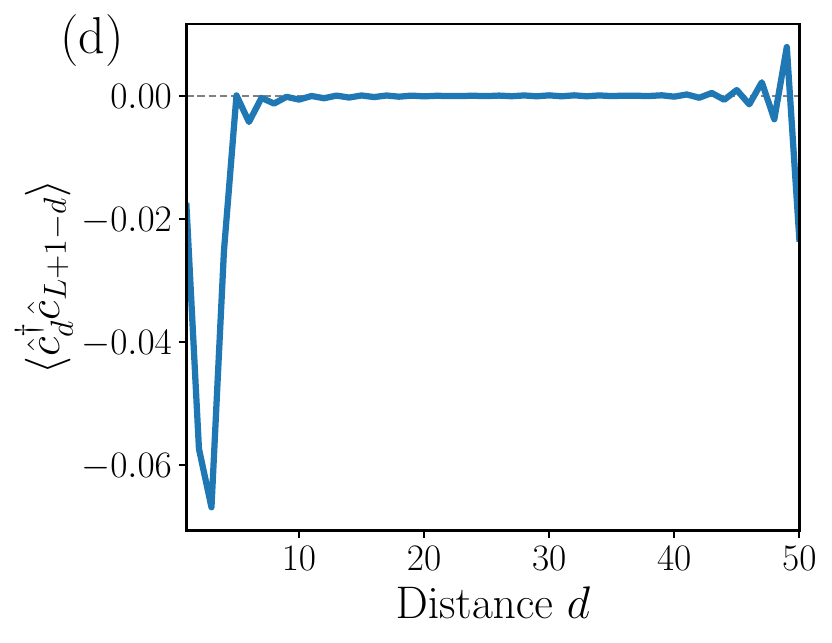}\label{fig:horseshoe}}
    \end{minipage}
    \caption{\protect\subref{fig:processes} Schematic representation of the model in Eq. \eqref{model}. The model describes spin-polarized fermions hopping on a one-dimensional lattice with nearest-neighbor attractive interactions. \protect\subref{fig:revival} Two-point correlator $\expvalue{\hat{c}^\dagger_1 \hat{c}_j}$ evaluated on the ground state of the model in Eq. \eqref{model} at half-filling $N=50, L=100$ for strong attractive interactions $V=-1.9$. The correlator goes to zero as $j$ is moved into the bulk, but revives to a finite value when $j$ approaches the opposite edge of the chain. \protect\subref{fig:horseshoe_setup} Schematic representation of the proposed experimental setup to measure the revival using beam-splitter interferometry. The spin-polarized fermions can hop along a one-dimensional horseshoe (highlighted in grey), since the red sites are blocked by a strong repulsive potential $U$. The labels $1, \dots, L$ of the sites of the chain are written in small black boxes. Performing beam-splitter interferometry between sites $d$ and $L+1-d$ allows us to measure the two-point correlator $\expvalue{\hat{c}^\dagger_d \hat{c}_{L+1-d}}$. \protect\subref{fig:horseshoe} Two-point correlator $\expvalue{\hat{c}^\dagger_d \hat{c}_{L+1-d}}$ evaluated on the ground state of the model in Eq. \eqref{model} at approximately half-filling $N=50, L=101$ for strong attractive interactions $V=-1.9$. The peak of the revival is located at $d=3$, which corresponds to a non-zero correlator between two points located at the opposite edges of the chain.}
\end{figure}

\emph{Introduction. ---} Majorana fermions are emergent zero-energy modes which are localized at the two edges of a one-dimensional topological superconductor \cite{kitaev_unpaired_2001,leijnse_introduction_2012}. Joined together, these two Majorana modes form a single fermionic degree of freedom which is highly non-local and thus robust against local perturbations and decoherence \cite{nayak_non-abelian_2008}. This robustness makes Majorana fermions a particularly promising platform for topological quantum computation and has motivated an intense search for their realization in both solid-state \cite{lutchyn_majorana_2010,oreg_helical_2010,alicea_new_2012} and ultracold atomic systems \cite{jiang_majorana_2011}.

The paradigmatic model hosting Majorana fermions is the Kitaev chain, a one-dimensional $p$-wave superconductor of spinless fermions \cite{kitaev_unpaired_2001}. The existence of Majorana modes in the Kitaev chain relies on two key ingredients: the model is not number-conserving and it possesses a bulk gap which protects the zero modes energetically and justifies the BCS mean-field treatment of the spectrum. In realistic quantum-simulation platforms, however, the microscopic dynamics usually conserves the total number of particles, and fermions in one-dimensional systems form a gapless Luttinger liquid, unless the interactions are specifically shaped \cite{giamarchi_quantum_2003}. Whether -- and in what form -- Majorana physics survives in such conditions is thus a question that is both conceptually and experimentally relevant, recently addressed in number-conserving models \cite{keselman_gapless_2015,iemini_majorana_2017,ruhman_topological_2017,fazzini_interaction-induced_2019,thomas-markarian_majorana_2025,bellinato_giacomelli_topology_2026} and in coupled wires \cite{fidkowski_majorana_2011,cheng_majorana_2011,sau_number_2011,kraus_majorana_2013,iemini_localized_2015,lang_topological_2015,liu_phase_2019,lisandrini_majorana_2022,tausendpfund_majorana_2023,defossez_dynamic_2025}.

In this work, we show that signatures of Majorana edge physics persist even in the conceptually simplest number-conserving, gapless Luttinger liquid of spinless fermions with nearest-neighbor attractive interactions (Fig.~\ref{fig:processes}):
\begin{align}
    \label{model}
    \hat{H} = -t\sum_{i=1}^{L} \left(\hat{c}_i^\dagger \hat{c}_{i+1} + \text{h.c.}\right) + V \sum_{i=1}^{L} \hat{n}_i \hat{n}_{i+1},
\end{align}
where $\hat{c}_i^\dagger$, $\hat{c}_i$ are fermionic creation and annihilation operators on site $i$, and $\hat{n}_i = \hat{c}_i^\dagger \hat{c}_i$ is the number operator. From now on, we set the hopping amplitude $t=1$ as the unit of energy. This model is a gapless Luttinger liquid for $V>-2$ and undergoes a phase transition to a phase-separated state for $V<-2$ \cite{giamarchi_quantum_2003}.

In the Luttinger liquid phase with strong attractive interactions, we show that the two-point correlator $\expvalue{\hat{c}^\dagger_1 \hat{c}_j}$ evaluated on the ground state of the system decays to zero as $j$ is moved into the bulk of the system, but revives to a finite value when $j$ approaches the opposite edge of the chain (Fig.~\ref{fig:revival}). The sign of the revival depends on the parity of the total number of fermions in the system. Additionally, we demonstrate that the revival is a robust feature that survives the thermodynamic limit, and persists at different fillings and in the low-lying excited states. Furthermore, a particularly simple ansatz proposed in \cite{thomas-markarian_majorana_2025} relates the odd and even ground states for a wide range of interaction strengths $V$.

Finally, we propose an experimental realization of the model in Eq. \eqref{model} using ultracold dipolar atoms or molecules in an optical lattice \cite{trefzger_ultracold_2011,baier_extended_2016,rosenberg_observation_2022,christakis_probing_2023,su_dipolar_2023,carroll_observation_2025}. The dipoles can hop along a one-dimensional horseshoe-shaped chain (Fig.~\ref{fig:horseshoe_setup}), and the two-point correlator $\expvalue{\hat{c}^\dagger_d \hat{c}_{L+1-d}}$ can be measured using beam-splitter interferometry between sites $d$ and $L+1-d$. This correlator exhibits a peak when $d$ corresponds to two points located at the opposite edges of the chain (Fig.~\ref{fig:horseshoe}), which is a smoking-gun signature of the non-local nature of the underlying Majorana modes present in the system.

\emph{Edge-to-edge revival. ---}
\begin{figure}[ht]
    \centering
    \captionsetup[subfigure]{labelformat=empty}
    \subfloat{\includegraphics[width=0.5\columnwidth, valign=c]{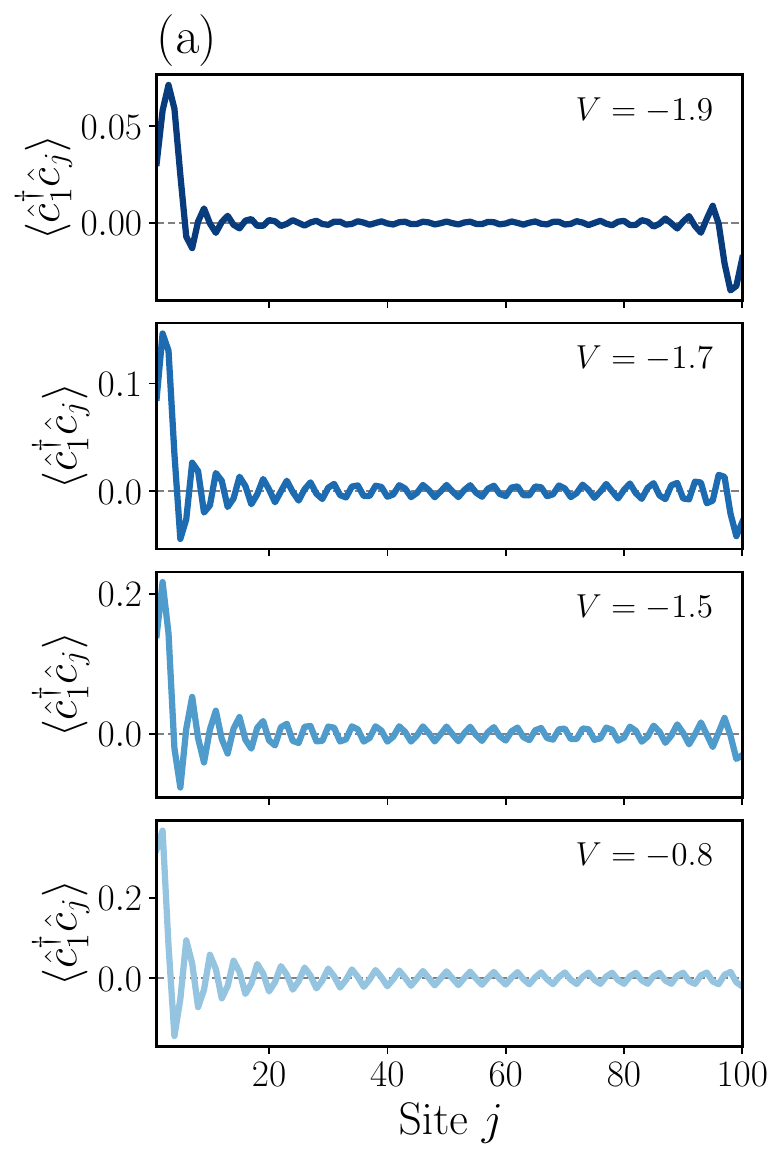}\label{fig:revival_V}}
    \hfill
    \subfloat{\includegraphics[width=0.5\columnwidth, valign=c]{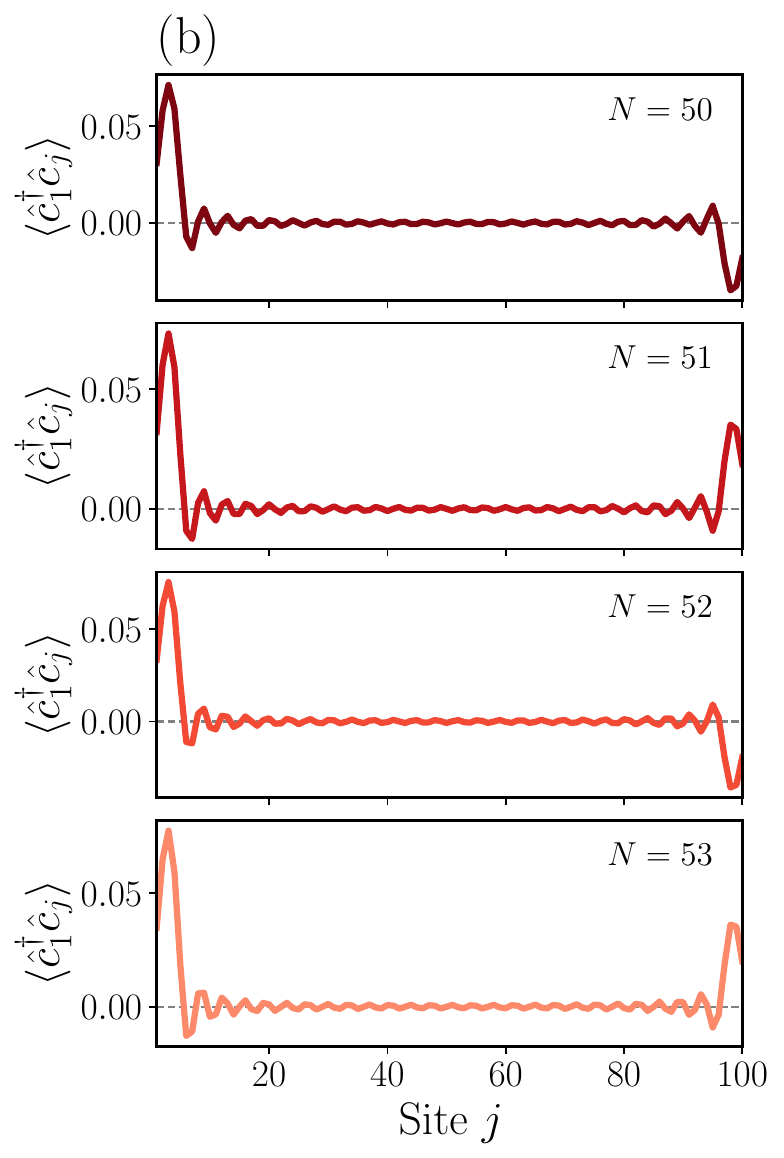}\label{fig:revival_N}}
    \caption{\protect\subref{fig:revival_V} Two-point correlator $\expvalue{\hat{c}^\dagger_1 \hat{c}_{j}}$ evaluated on the ground state of the model in Eq. \eqref{model} at half-filling $N=50, L=100$ for different values of the interaction strength $V$. The correlator goes to zero as $j$ is moved into the bulk, but revives to a finite value when $j$ is located at the opposite edge of the chain and $V$ is sufficiently large. \protect\subref{fig:revival_N} Two-point correlator $\expvalue{\hat{c}^\dagger_1 \hat{c}_{j}}$ evaluated on the ground state of the model in Eq. \eqref{model} at strong interaction strength $V=-1.9$ and $L=100$, for different values of the total fermion number $N$. The revival is robust to adding fermions to the half-filled chain, but its sign depends on the parity of the number of fermions $N$ in the system.}
\end{figure}
In the Kitaev chain, Majorana fermions correspond to two zero-energy modes localized at the left and right edges of the chain \cite{kitaev_unpaired_2001}. At finite sizes, these two zero modes hybridize and acquire a finite energy splitting \cite{cheng_splitting_2009}. The localization of these two hybridized zero modes can be characterized by the two-point correlator $\expvalue{\hat{c}_i^\dagger \hat{c}_j}$ evaluated on the ground state of the system. For the Kitaev chain, the two-point correlator decays exponentially with the distance $|i-j|$ between the two sites when the first point $i$ is located in the bulk. The presence of modes localized at the edges of the chain is signaled by a revival of the two-point correlator when the first point $i=1$ is located at one edge of the system. The correlator $\expvalue{\hat{c}^\dagger_1 \hat{c}_{j}}$ goes to zero as $j$ is moved into the bulk of the system, but revives to a finite value when $j$ is located at the opposite edge of the chain. The revival is thus a robust fingerprint of the underlying Majorana modes present in the Kitaev model.

In Fig.~\ref{fig:revival_V}, we show that the two-point correlator $\expvalue{\hat{c}^\dagger_1 \hat{c}_{j}}$ evaluated on the ground state of the model in Eq. \eqref{model} also exhibits a revival when $j$ approaches the opposite edge of the chain, provided the attractive interactions are strong enough. At the same time, $\expvalue{\hat{c}^\dagger_{i} \hat{c}_{j}} \propto |i-j|^{-\alpha}$ decays as a power-law in the bulk \cite{giamarchi_quantum_2003}, with an exponent $\alpha=(K+1/K)/2$ that depends on the Luttinger parameter $K$ of the system. 

As a function of the interaction strength $V$, the revival emerges more distinctly for values of $V$ which approach the critical value $V_c=-2$ from above. For weak attractive interactions, the correlator approaches the free-fermion regime, for which the correlator in the bulk of the system is finite and oscillates with a period of $k_F$ without reviving at the opposite end of the chain (see lowest panel of Fig.~\ref{fig:revival_V} for $V=-0.8$). For stronger attractive interactions, the fermions' tendency to pair is enhanced and the correlator becomes more suppressed in the bulk of the system. This reflects the divergence of the Luttinger liquid parameter, $K\rightarrow \infty$, as $V$ approaches $V_c$, leading to bulk correlations $\expvalue{\hat{c}^\dagger_{i} \hat{c}_{j}} \propto |i-j|^{-K/2}$  becoming arbitrarily short-ranged close to $V_c=-2$.

The model in Eq. \eqref{model} conserves the total number of fermions $N$. In Fig.~\ref{fig:revival_N}, we show the dependence of the revival on the number of fermions $N$. In particular, we show that the magnitude of the revival is robust to adding fermions to the half-filled chain. The sign of the revival, however, depends on the parity of the number of fermions $N$ in the system. This constitutes a direct manifestation of the underlying Majorana modes present in the system, which are known to be sensitive to the parity of the number of fermions.

This can be understood directly for the Kitaev chain at the special point $t=\Delta$ and $\mu=0$, for which the two Majorana modes are exactly localized at the edges of the chain and they can be expressed in terms of the original fermionic operators $\hat{c}_i$ as $\hat{\gamma}_L = \hat{c}_1^\dagger + \hat{c}_1$ and $\hat{\gamma}_R = i(\hat{c}_L^\dagger - \hat{c}_L)$. The two Majorana operators $\hat{\gamma}_L$ and $\hat{\gamma}_R$ can be used to build a non-local fermionic operator $\hat{d}^\dagger = \hat{\gamma}_L - i \hat{\gamma}_R$, which is an exact zero-energy quasiparticle operator. For this reason, the ground state of the Kitaev chain is two-fold degenerate, and the two ground states $\ket{\text{even}}$ and $\ket{\text{odd}}$ differ by the parity of the number of fermions. In particular, $\hat{d}\ket{\text{even}} = 0$ and the $\hat{d}^\dagger$ quasiparticle operator connects the two ground states with opposite parities
\begin{align}
    \label{eq:two_ground_states_kitaev}
    \hat{d}^\dagger \ket{\text{even}} = \ket{\text{odd}}.
\end{align}
The parity operator $\hat{P}$ in the ground-state subspace is thus
\begin{align}
    \label{eq:parity_correlator_kitaev}
    \hat{P} = 1-2\hat{d}^\dagger \hat{d}=-i\hat{\gamma}_L\hat{\gamma}_R = -(\hat{c}_1^\dagger \hat{c}_L+\hat{c}_1 \hat{c}_L+\text{h.c.}).
\end{align}
Eq. \eqref{eq:parity_correlator_kitaev} shows that in the Kitaev chain the parity of the ground state $\expvalue{\hat{P}}$ is directly connected to the expectation values of the non-local normal and anomalous two-point correlators $\expvalue{\hat{c}_1^\dagger \hat{c}_L}$ and $\expvalue{\hat{c}_1 \hat{c}_L}$.
In number-conserving models, such as the one in Eq. \eqref{model}, the anomalous correlator is zero and the sign of the revival $\expvalue{\hat{c}_1^\dagger \hat{c}_L}$ directly dictates the parity of the ground state (see Fig.~\ref{fig:revival_N}). More specifically, in the number-conserving case it is possible to build two self-adjoint operators $\hat{\gamma}_L$ and $\hat{\gamma}_R$ (localized at the left and right edges of the chain, respectively) such that the parity $P(N)$ of the ground state $\ket{N}$ with $N$ fermions is given by $P(N)=-ic\ev{\hat{\gamma}_L\hat{\gamma}_R}{N}$, where $c$ is some $N$-independent constant \cite{thomas-markarian_majorana_2025}.

It is important to note that the revival is not a finite-size effect (see Supplemental Material for a detailed numerical analysis of the scaling of the revival). In analogy to the Kitaev chain, the revival is a manifestation of proper, normalized quasiparticle operators, which are unaffected by the $L\to\infty$ limit. Furthermore, the revival is not a peculiar feature of the ground state of the system at half-filling. In fact, we show that the revival is also present at low fillings with a spatial extension which is comparable to the Fermi wavelength of the system and is thus filling-dependent (see Supplemental Material for a discussion of the revival at low and high fillings). Finally, we find that the revival is also present in the low-lying excited states of the system (see Supplemental Material for an analytical treatment using bosonization of the two-point correlator evaluated on the low-lying excited states). In a Luttinger liquid, the low-energy excited states emerge on top of the ground state as long-wavelength density fluctuations, which do not affect the short-wavelength physics of the revival.

\emph{Interpolating from Majorana to free fermions. ---}
\begin{figure}[ht]
    \centering
    \captionsetup[subfigure]{labelformat=empty}
    \subfloat{\hspace{0.5em}\includegraphics[width=\columnwidth, valign=c]{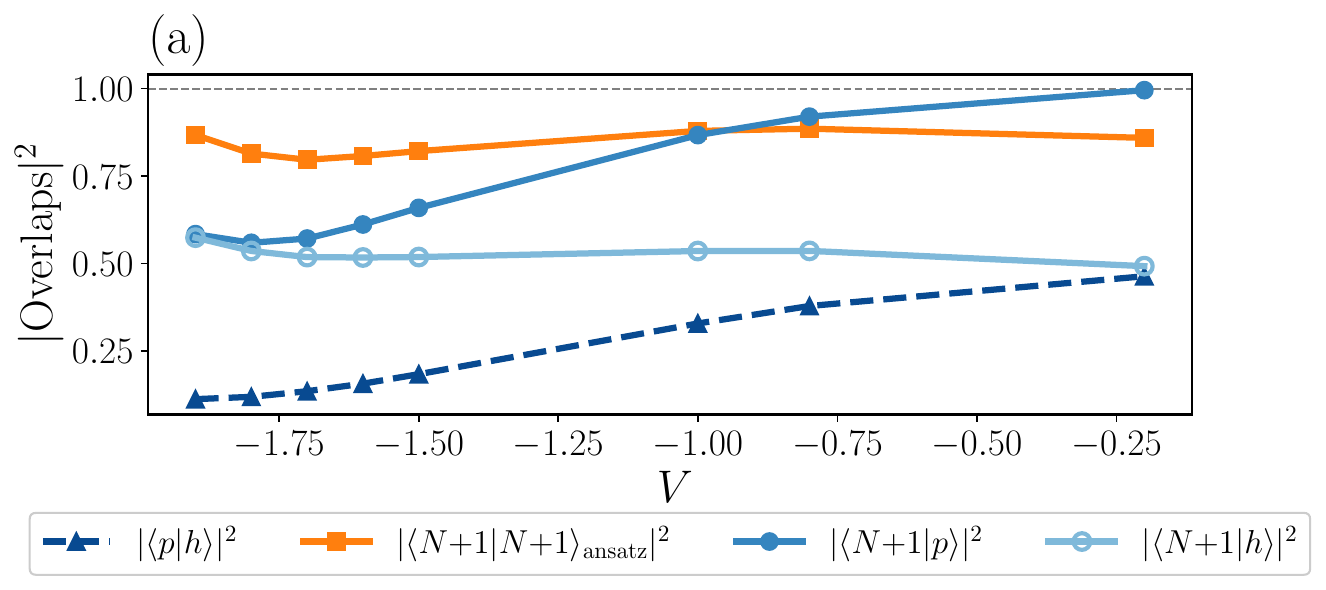}\label{fig:overlaps}}
    \vspace{-1.5em}
    \subfloat{\includegraphics[width=\columnwidth, valign=c]{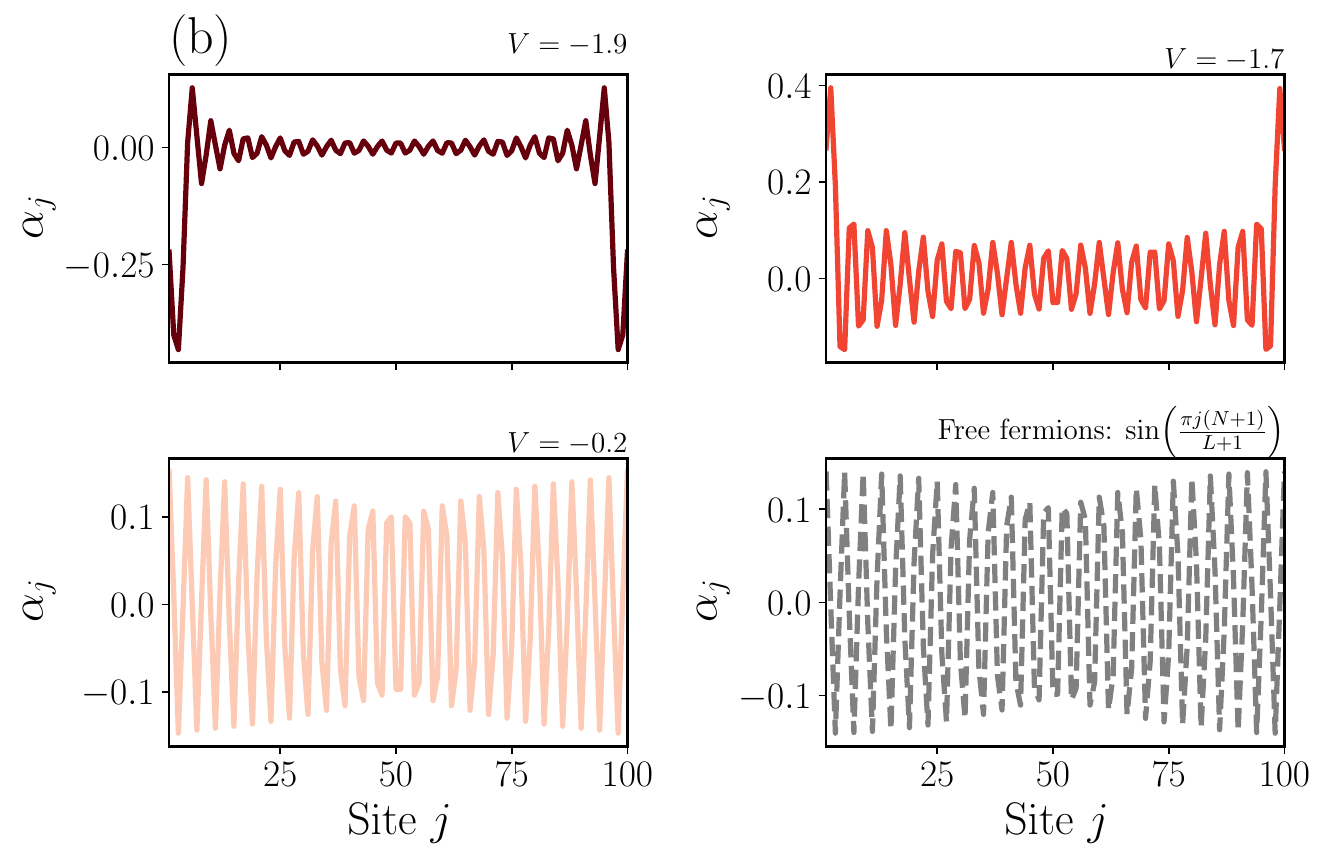}\label{fig:alphas}}
    \caption{\protect\subref{fig:overlaps} Squared overlap $|\bra{N+1}\ket{N+1}_{\text{ansatz}}|^2$ of the ansatz in Eq. \eqref{ansatz} with the exact ground state $\ket{N+1}$ of the system with $N+1$ fermions ($N=50, L=100$). The squared overlap is very high $|\bra{N+1}\ket{N+1}_{\text{ansatz}}|^2 \approx 0.85$ for a wide range of interaction strengths $V$. \protect\subref{fig:alphas} Coefficients $\alpha_i$ used in the ansatz in Eq. \eqref{ansatz} as a function of interaction strength $V$. The coefficients $\alpha_i$ are localized at the edges of the chain for strong attractive interactions, while they are delocalized in the bulk of the system for weak attractive interactions, smoothly interpolating between the Majorana and free-fermion limits. The analytical expression for the coefficients $\alpha_i$ in the free-fermion limit $V=0$ is displayed as a grey dashed line.}
\end{figure}
A key signature of the presence of Majorana modes in gapped models that do not conserve particle number, such as the Kitaev chain, is the two-fold degeneracy of the ground state. In particular, the two ground states differ by the parity of the number of fermions, and they are related by the action of a quasiparticle fermionic operator $\hat{d}^\dagger$ (see Eq. \eqref{eq:two_ground_states_kitaev}). In terms of the original fermionic operators $\hat{c}_i$, one can express the quasiparticle operator as
\begin{align}
    \label{eq:quasiparticle_operator}
    \hat{d}^\dagger = \frac{1}{\sqrt{2}}\sum_{j=1}^{L}\left(\alpha_j \hat{c}_j^\dagger + \beta_j \hat{c}_j\right)
\end{align}
where the coefficients $\alpha_j$ and $\beta_j$ are exponentially localized at the edges of the chain and form two orthonormal vectors.

In number-conserving models, such as the one in Eq. \eqref{model}, a similar construction that relates the even- and odd-parity ground states can be made \cite{thomas-markarian_majorana_2025}. In particular, one can make the following ansatz for the ground state $\ket{N+1}$ with an odd number of fermions $N+1$:
\begin{align}
    \label{ansatz}
    \ket{N+1}_{\text{ansatz}} \propto \sum_{j=1}^L \alpha_j \hat{c}_j^\dagger \ket{N} + \sum_{j=1}^{L} \beta_j \hat{c}_j \ket{N+2}.
\end{align}
The $\alpha_i$ and $\beta_i$ vectors are the two orthonormal eigenvectors corresponding to the only non-zero eigenvalues of the matrix $\Delta C_{ij} = \bra{N+1} \hat{c}_i^\dagger \hat{c}_j \ket{N+1} - \bra{N} \hat{c}_i^\dagger \hat{c}_j \ket{N}$, which can be shown numerically to have approximately rank two \cite{thomas-markarian_majorana_2025}.

As shown in the left panel of Fig.~\ref{fig:overlaps}, the ansatz in Eq. \eqref{ansatz} has a very high squared overlap with the exact ground state of the system $|\bra{N+1}\ket{N+1}_{\text{ansatz}}|^2 \approx 0.85$ for a wide range of interaction strengths $V$. To explain this, we decompose the ansatz in Eq. \eqref{ansatz} into two parts $\ket{N+1}_{\text{ansatz}} \propto \ket{p} + \ket{h}$, where $\ket{p} \propto \sum_{j=1}^L \alpha_j \hat{c}_j^\dagger \ket{N}$ and $\ket{h} \propto \sum_{j=1}^{L} \beta_j \hat{c}_j \ket{N+2}$ are the particle and hole components, respectively.

In the limit of strong attractive interactions $V = -1.9$, the particle and hole components are almost orthogonal $|\bra{p} \ket{h}|^2 \approx 0.1$, and the ground state is well approximated by an equal-weight superposition of the two components $|\bra{N+1}\ket{p}|^2 \approx |\bra{N+1}\ket{h}|^2 \approx 0.6$. Furthermore, the $\alpha_i$ and $\beta_i$ coefficients are localized at the edges of the chain (see Fig.~\ref{fig:alphas}). This is analogous to the quasiparticle operator $\hat{d}^\dagger$ introduced for the Kitaev chain in Eq. \eqref{eq:quasiparticle_operator}, which is an equal-weight superposition of a particle and a hole component.

This picture changes drastically as the interaction strength is decreased, but the ansatz in Eq. \eqref{ansatz} still captures the ground state of the system with a high overlap. For $V = -0.2$, the ground state is well approximated by the particle component alone $|\bra{N+1}\ket{p}|^2 \approx 1$, and the $\alpha_i$ coefficients are delocalized in the bulk of the system. This can be understood explicitly in the free-fermion limit $V=0$ in an open chain, where the ground state of the system with an odd number of fermions $N+1$ is obtained by adding a fermion with momentum $k_F$ to the ground state of the system with an even number of fermions $N$. In other words, $\ket{N+1} = \hat{c}_{k_F}^\dagger \ket{N}$, where $\hat{c}_{k_F}^\dagger = \sqrt{2/(L+1)} \sum_{j=1}^L \sin\left(k_F j\right) \hat{c}_j^\dagger$ and $k_F = \pi (N+1)/(L+1)$ is the Fermi momentum.

As shown in the right panel of Fig.~\ref{fig:overlaps}, the $\alpha_i$ coefficients interpolate smoothly between the Majorana-fermion regime for strong attractive interactions, in which the coefficients are localized at the edges of the chain, and the free-fermion regime for weak attractive interactions, in which the coefficients are delocalized in the bulk of the system. For this reason, the ansatz in Eq. \eqref{ansatz} is a very good approximation of the ground state of the system for a wide range of interaction strengths $V$.

\emph{Experimental realization and outlook. ---} 
The strong, controllable attractive interactions required to observe the revival in the model in Eq. \eqref{model} can be realized with ultracold dipolar quantum gases in an optical lattice. We checked numerically that all the results obtained in this work also apply to the case of dipolar $1/r^3$ interactions, since they are essentially short-ranged in one dimension \cite{giamarchi_quantum_2003}. Two routes are particularly promising: ultracold molecules, whose dipole-dipole interaction is dressed by a microwave field \cite{cooper_stable_2009,karman_microwave_2018,chen_field-linked_2023,bigagli_observation_2024}, giving direct control over both the sign and the strength of the interaction; or atoms with an induced \cite{guardado-sanchez_quench_2021} or permanent dipole moment \cite{lu_quantum_2012,aikawa_observation_2014,chomaz_dipolar_2022}.

To detect the revival, we propose a single-chain interferometric protocol. The molecules or atoms are loaded in a horseshoe-shaped chain (see Fig.~\ref{fig:horseshoe_setup}) and the ground state of the system is prepared. A beam-splitter operation is then performed between sites $d$ and $L+1-d$, which maps the $\hat{c}_d$ operator to the new operator $\hat{c}'_d$:
\begin{align}
    \hat{c}_d \rightarrow \hat{c}'_d \equiv \frac{1}{\sqrt{2}}(\hat{c}_d+ \hat{c}_{L+1-d}).
\end{align}
Measuring the occupation number at site $d$ after the beam-splitter operation is equivalent to measuring the number operator
\begin{align}
    \hat{n}'_d = \hat{c}_d'^\dagger \hat{c}'_d = \frac{1}{2}(\hat{n}_d + \hat{n}_{L+1-d})+\frac{1}{2}(\hat{c}_d^\dagger \hat{c}_{L+1-d}+\text{h.c.}),
\end{align}
where $\hat{n}_i = \hat{c}_i^\dagger \hat{c}_i$. This protocol, which consists of beam-splitter operations and single-site occupancy measurements, thus allows us to directly reconstruct the two-point correlator $\expvalue{\hat{c}_d^\dagger \hat{c}_{L+1-d}}$ \cite{impertro_strongly_2025} and to detect the revival in the system (see Fig.~\ref{fig:horseshoe}). Crucially, since the revival survives in the low-lying excited states of the system, a thermal mixture of low-energy states is sufficient to observe the revival, thus relaxing the cooling requirements for the experiment. Furthermore, an experimentally relevant quantity is given by the ratio of the revival's magnitude to the bulk value of the two-point correlator $|\expvalue{\hat{c}_1^\dagger \hat{c}_{L/2}}|$. Since we are in the Luttinger liquid phase, the bulk value of the two-point correlator near the phase transition $V \rightarrow V_c=-2$ is suppressed as
\begin{align}
    \label{eq:suppression}
    |\expvalue{\hat{c}_1^\dagger \hat{c}_{L/2}}| \propto L^{-K/2} = L^{-1/\sqrt{\epsilon}},
\end{align}
where $\epsilon \propto 2+V$ at half-filling \cite{giamarchi_quantum_2003}. Remarkably, even in finite-size systems, the two-point bulk correlator $|\expvalue{\hat{c}_1^\dagger \hat{c}_{L/2}}|$ can be heavily suppressed by fine-tuning $V$ close to the critical point $V_c$. This highlights a major advantage of cold-atom platforms, where the interaction strength $V$ serves as a highly adjustable experimental tuning knob.

An important direction for future work is to understand whether these Majorana modes in spinless Luttinger liquids can be stabilized and protected to realize a topological qubit. Coupling two copies of the model in Eq. \eqref{model} to a common third wire, in a spirit analogous to Ref. \cite{fidkowski_majorana_2011}, could similarly pin the edge modes identified in this work. Notably, the specific scaling behavior near the phase transition in Eq. \eqref{eq:suppression} plays a beneficial role by efficiently suppressing single-particle excitations and could provide an advantage in turning this number-conserving, cold-atom-compatible Luttinger liquid into a genuinely topologically protected qubit.

\emph{Methods. ---} All the numerical results presented in this work were obtained using DMRG simulations performed with the SyTen library \cite{hubig_syten_nodate}.

\emph{Acknowledgments. ---} The authors thank Luca Muscarella, Chengfeng Xu and Xin-Yu Luo for insightful discussions on the experimental realization of the model. F.D. acknowledges support from the International Max Planck Research School. N.C. acknowledges ISCRA for awarding this project access to the LEONARDO supercomputer, owned by the EuroHPC Joint Undertaking, hosted by CINECA (Italy) and acknowledges financial support from the ICSC - Centro Nazionale di Ricerca in High Performance Computing, Big Data and Quantum Computing, funded by European Union - NextGenerationEU (Grant number CN00000013). L. B. acknowledges financial support within the DiQut Grant No.2022523NA7 funded by European Union - Next Generation EU, PRIN 2022 program (D.D. 104 - 02/02/2022 Ministero dell'Università e della Ricerca). This project has received funding from the Deutsche Forschungsgemeinschaft (DFG, German Research Foundation) under Germany's Excellence Strategy -- EXC-2111 -- 390814868, and from the European Research Council (ERC) under the European Union's Horizon 2020 research and innovation programme (Grant Agreement no 948141) — ERC Starting Grant SimUcQuam.

\bibliographystyle{unsrturl}
\bibliography{1d_majoranas.bib}

@article{kitaev_unpaired_2001,
	title = {Unpaired {Majorana} fermions in quantum wires},
	volume = {44},
	issn = {1468-4780},
	url = {http://arxiv.org/abs/cond-mat/0010440},
	doi = {10.1070/1063-7869/44/10S/S29},
	language = {en},
	number = {10S},
	urldate = {2026-03-25},
	journal = {Physics-Uspekhi},
	author = {Kitaev, Alexei},
	month = oct,
	year = {2001},
	note = {arXiv:cond-mat/0010440},
	pages = {131--136},
}

@article{cheng_splitting_2009,
	title = {Splitting of {Majorana} modes due to intervortex tunneling in a p + ip superconductor},
	volume = {103},
	issn = {0031-9007, 1079-7114},
	url = {http://arxiv.org/abs/0905.0035},
	doi = {10.1103/PhysRevLett.103.107001},
	language = {en},
	number = {10},
	urldate = {2026-03-25},
	journal = {Physical Review Letters},
	author = {Cheng, Meng and Lutchyn, Roman M. and Galitski, Victor and Sarma, S. Das},
	month = aug,
	year = {2009},
	note = {arXiv:0905.0035 [cond-mat]},
	pages = {107001},
}

@article{leijnse_introduction_2012,
	title = {Introduction to topological superconductivity and {Majorana} fermions},
	volume = {27},
	issn = {0268-1242, 1361-6641},
	url = {http://arxiv.org/abs/1206.1736},
	doi = {10.1088/0268-1242/27/12/124003},
	language = {en},
	number = {12},
	urldate = {2026-03-25},
	journal = {Semiconductor Science and Technology},
	author = {Leijnse, Martin and Flensberg, Karsten},
	month = dec,
	year = {2012},
	note = {arXiv:1206.1736 [cond-mat]},
	pages = {124003},
}

@book{giamarchi_quantum_2003,
	address = {Oxford, New York},
	series = {International {Series} of {Monographs} on {Physics}},
	title = {Quantum {Physics} in {One} {Dimension}},
	isbn = {978-0-19-852500-4},
	publisher = {Oxford University Press},
	author = {Giamarchi, Thierry},
	month = dec,
	year = {2003},
}

@misc{thomas-markarian_majorana_2025,
	title = {Majorana edge modes in number-conserving models with long-range interactions},
	url = {http://arxiv.org/abs/2509.00158},
	doi = {10.48550/arXiv.2509.00158},
	language = {en},
	urldate = {2026-03-26},
	publisher = {arXiv},
	author = {Thomas-Markarian, Jaden and Agarwal, Kartiek and Martin, Ivar},
	month = aug,
	year = {2025},
	note = {arXiv:2509.00158 [cond-mat]},
}

@article{nayak_non-abelian_2008,
	title = {Non-{Abelian} anyons and topological quantum computation},
	volume = {80},
	copyright = {http://link.aps.org/licenses/aps-default-license},
	issn = {0034-6861, 1539-0756},
	url = {https://link.aps.org/doi/10.1103/RevModPhys.80.1083},
	doi = {10.1103/RevModPhys.80.1083},
	language = {en},
	number = {3},
	urldate = {2026-04-03},
	journal = {Reviews of Modern Physics},
	author = {Nayak, Chetan and Simon, Steven H. and Stern, Ady and Freedman, Michael and Das Sarma, Sankar},
	month = sep,
	year = {2008},
	pages = {1083--1159},
}

@article{ruhman_topological_2017,
	title = {Topological degeneracy and pairing in a one-dimensional gas of spinless {Fermions}},
	volume = {96},
	issn = {2469-9950, 2469-9969},
	url = {http://arxiv.org/abs/1705.00635},
	doi = {10.1103/PhysRevB.96.085133},
	language = {en},
	number = {8},
	urldate = {2026-05-15},
	journal = {Physical Review B},
	author = {Ruhman, Jonathan and Altman, Ehud},
	month = aug,
	year = {2017},
	note = {arXiv:1705.00635 [cond-mat.quant-gas]},
	pages = {085133},
}

@misc{hubig_syten_nodate,
	title = {The {SyTen} {Toolkit}},
	url = {https://syten.eu},
	author = {Hubig, Claudius and Lachenmaier, Felix and Linden, Nils-Oliver and Reinhard, Teresa and Stenzel, Leo and Swoboda, Andreas and Grundner, Martin and Mardazad, Sam and Paeckel, Sebastian},
}

@article{su_dipolar_2023,
	title = {Dipolar quantum solids emerging in a {Hubbard} quantum simulator},
	volume = {622},
	copyright = {2023 The Author(s), under exclusive licence to Springer Nature Limited},
	issn = {1476-4687},
	url = {https://www.nature.com/articles/s41586-023-06614-3},
	doi = {10.1038/s41586-023-06614-3},
	language = {en},
	number = {7984},
	urldate = {2026-07-02},
	journal = {Nature},
	publisher = {Nature Publishing Group},
	author = {Su, Lin and Douglas, Alexander and Szurek, Michal and Groth, Robin and Ozturk, S. Furkan and Krahn, Aaron and Hébert, Anne H. and Phelps, Gregory A. and Ebadi, Sepehr and Dickerson, Susannah and Ferlaino, Francesca and Marković, Ognjen and Greiner, Markus},
	month = oct,
	year = {2023},
	pages = {724--729},
}

@article{impertro_strongly_2025,
	title = {Strongly interacting {Meissner} phases in large bosonic flux ladders},
	volume = {21},
	copyright = {2025 The Author(s)},
	issn = {1745-2481},
	url = {https://www.nature.com/articles/s41567-025-02890-0},
	doi = {10.1038/s41567-025-02890-0},
	language = {en},
	number = {6},
	urldate = {2026-07-02},
	journal = {Nature Physics},
	publisher = {Nature Publishing Group},
	author = {Impertro, Alexander and Huh, SeungJung and Karch, Simon and Wienand, Julian F. and Bloch, Immanuel and Aidelsburger, Monika},
	month = jun,
	year = {2025},
	pages = {895--901},
}

@article{chen_field-linked_2023,
	title = {Field-linked resonances of polar molecules},
	volume = {614},
	copyright = {2023 The Author(s)},
	issn = {1476-4687},
	url = {https://www.nature.com/articles/s41586-022-05651-8},
	doi = {10.1038/s41586-022-05651-8},
	language = {en},
	number = {7946},
	urldate = {2026-07-02},
	journal = {Nature},
	publisher = {Nature Publishing Group},
	author = {Chen, Xing-Yan and Schindewolf, Andreas and Eppelt, Sebastian and Bause, Roman and Duda, Marcel and Biswas, Shrestha and Karman, Tijs and Hilker, Timon and Bloch, Immanuel and Luo, Xin-Yu},
	month = feb,
	year = {2023},
	pages = {59--63},
}

@article{defossez_dynamic_2025,
	title = {Dynamic realization of {Majorana} zero modes in a particle-conserving ladder},
	volume = {7},
	issn = {2643-1564},
	url = {https://link.aps.org/doi/10.1103/PhysRevResearch.7.023183},
	doi = {10.1103/PhysRevResearch.7.023183},
	language = {en},
	number = {2},
	urldate = {2026-07-06},
	journal = {Physical Review Research},
	author = {Defossez, Anaïs and Vanderstraeten, Laurens and Peralta Gavensky, Lucila and Goldman, Nathan},
	month = may,
	year = {2025},
	pages = {023183},
}

@article{cheng_majorana_2011,
	title = {Majorana edge states in interacting two-chain ladders of fermions},
	volume = {84},
	copyright = {http://link.aps.org/licenses/aps-default-license},
	issn = {1098-0121, 1550-235X},
	url = {https://link.aps.org/doi/10.1103/PhysRevB.84.094503},
	doi = {10.1103/PhysRevB.84.094503},
	language = {en},
	number = {9},
	urldate = {2026-07-06},
	journal = {Physical Review B},
	author = {Cheng, Meng and Tu, Hong-Hao},
	month = sep,
	year = {2011},
	pages = {094503},
}

@article{tausendpfund_majorana_2023,
	title = {Majorana zero modes in fermionic wires coupled by {Aharonov}-{Bohm} cages},
	volume = {107},
	issn = {2469-9950, 2469-9969},
	url = {https://link.aps.org/doi/10.1103/PhysRevB.107.035124},
	doi = {10.1103/PhysRevB.107.035124},
	language = {en},
	number = {3},
	urldate = {2026-07-06},
	journal = {Physical Review B},
	author = {Tausendpfund, Niklas and Diehl, Sebastian and Rizzi, Matteo},
	month = jan,
	year = {2023},
	pages = {035124},
}

@article{sau_number_2011,
	title = {Number conserving theory for topologically protected degeneracy in one-dimensional fermions},
	volume = {84},
	copyright = {http://link.aps.org/licenses/aps-default-license},
	issn = {1098-0121, 1550-235X},
	url = {https://link.aps.org/doi/10.1103/PhysRevB.84.144509},
	doi = {10.1103/PhysRevB.84.144509},
	language = {en},
	number = {14},
	urldate = {2026-07-06},
	journal = {Physical Review B},
	author = {Sau, Jay D. and Halperin, B. I. and Flensberg, K. and Das Sarma, S.},
	month = oct,
	year = {2011},
	pages = {144509},
}

@article{fidkowski_majorana_2011,
	title = {Majorana zero modes in one-dimensional quantum wires without long-ranged superconducting order},
	volume = {84},
	copyright = {http://link.aps.org/licenses/aps-default-license},
	issn = {1098-0121, 1550-235X},
	url = {https://link.aps.org/doi/10.1103/PhysRevB.84.195436},
	doi = {10.1103/PhysRevB.84.195436},
	language = {en},
	number = {19},
	urldate = {2026-07-06},
	journal = {Physical Review B},
	author = {Fidkowski, Lukasz and Lutchyn, Roman M. and Nayak, Chetan and Fisher, Matthew P. A.},
	month = nov,
	year = {2011},
	pages = {195436},
}

@article{rosenberg_observation_2022,
	title = {Observation of the {Hanbury} {Brown}–{Twiss} effect with ultracold molecules},
	volume = {18},
	copyright = {2022 The Author(s), under exclusive licence to Springer Nature Limited},
	issn = {1745-2481},
	url = {https://www.nature.com/articles/s41567-022-01695-9},
	doi = {10.1038/s41567-022-01695-9},
	language = {en},
	number = {9},
	urldate = {2026-07-07},
	journal = {Nature Physics},
	publisher = {Nature Publishing Group},
	author = {Rosenberg, Jason S. and Christakis, Lysander and Guardado-Sanchez, Elmer and Yan, Zoe Z. and Bakr, Waseem S.},
	month = sep,
	year = {2022},
	pages = {1062--1066},
}

@article{christakis_probing_2023,
	title = {Probing site-resolved correlations in a spin system of ultracold molecules},
	volume = {614},
	copyright = {2023 The Author(s), under exclusive licence to Springer Nature Limited},
	issn = {1476-4687},
	url = {https://www.nature.com/articles/s41586-022-05558-4},
	doi = {10.1038/s41586-022-05558-4},
	language = {en},
	number = {7946},
	urldate = {2026-07-07},
	journal = {Nature},
	publisher = {Nature Publishing Group},
	author = {Christakis, Lysander and Rosenberg, Jason S. and Raj, Ravin and Chi, Sungjae and Morningstar, Alan and Huse, David A. and Yan, Zoe Z. and Bakr, Waseem S.},
	month = feb,
	year = {2023},
	pages = {64--69},
}

@article{guardado-sanchez_quench_2021,
	title = {Quench {Dynamics} of a {Fermi} {Gas} with {Strong} {Nonlocal} {Interactions}},
	volume = {11},
	url = {https://link.aps.org/doi/10.1103/PhysRevX.11.021036},
	doi = {10.1103/PhysRevX.11.021036},
	number = {2},
	urldate = {2026-07-07},
	journal = {Physical Review X},
	publisher = {American Physical Society},
	author = {Guardado-Sanchez, Elmer and Spar, Benjamin M. and Schauss, Peter and Belyansky, Ron and Young, Jeremy T. and Bienias, Przemyslaw and Gorshkov, Alexey V. and Iadecola, Thomas and Bakr, Waseem S.},
	month = may,
	year = {2021},
	pages = {021036},
}

@article{lu_quantum_2012,
	title = {Quantum {Degenerate} {Dipolar} {Fermi} {Gas}},
	volume = {108},
	copyright = {http://link.aps.org/licenses/aps-default-license},
	issn = {0031-9007, 1079-7114},
	url = {https://link.aps.org/doi/10.1103/PhysRevLett.108.215301},
	doi = {10.1103/PhysRevLett.108.215301},
	language = {en},
	number = {21},
	urldate = {2026-07-07},
	journal = {Physical Review Letters},
	author = {Lu, Mingwu and Burdick, Nathaniel Q. and Lev, Benjamin L.},
	month = may,
	year = {2012},
	pages = {215301},
}

@article{aikawa_observation_2014,
	title = {Observation of {Fermi} surface deformation in a dipolar quantum gas},
	volume = {345},
	url = {https://www.science.org/doi/10.1126/science.1255259},
	doi = {10.1126/science.1255259},
	number = {6203},
	urldate = {2026-07-07},
	journal = {Science},
	publisher = {American Association for the Advancement of Science},
	author = {Aikawa, K. and Baier, S. and Frisch, A. and Mark, M. and Ravensbergen, C. and Ferlaino, F.},
	month = sep,
	year = {2014},
	pages = {1484--1487},
}

@article{kraus_majorana_2013,
	title = {Majorana {Edge} {States} in {Atomic} {Wires} {Coupled} by {Pair} {Hopping}},
	volume = {111},
	copyright = {http://link.aps.org/licenses/aps-default-license},
	issn = {0031-9007, 1079-7114},
	url = {https://link.aps.org/doi/10.1103/PhysRevLett.111.173004},
	doi = {10.1103/PhysRevLett.111.173004},
	language = {en},
	number = {17},
	urldate = {2026-07-07},
	journal = {Physical Review Letters},
	author = {Kraus, Christina V. and Dalmonte, Marcello and Baranov, Mikhail A. and Läuchli, Andreas M. and Zoller, P.},
	month = oct,
	year = {2013},
	pages = {173004},
}

@article{liu_phase_2019,
	title = {Phase diagram of interacting fermionic two-leg ladder with pair hopping},
	volume = {28},
	copyright = {http://iopscience.iop.org/info/page/text-and-data-mining},
	issn = {1674-1056, 2058-3834},
	url = {https://iopscience.iop.org/article/10.1088/1674-1056/28/2/020303},
	doi = {10.1088/1674-1056/28/2/020303},
	language = {en},
	number = {2},
	urldate = {2026-07-07},
	journal = {Chinese Physics B},
	author = {Liu, Wan-Li and Yuan, Tian-Zhong and Lin, Zhi and Yan, Wei},
	month = feb,
	year = {2019},
	pages = {020303},
}

@article{iemini_majorana_2017,
	title = {Majorana {Quasiparticles} {Protected} by {Z} 2 {Angular} {Momentum} {Conservation}},
	volume = {118},
	copyright = {http://link.aps.org/licenses/aps-default-license},
	issn = {0031-9007, 1079-7114},
	url = {http://link.aps.org/doi/10.1103/PhysRevLett.118.200404},
	doi = {10.1103/PhysRevLett.118.200404},
	language = {en},
	number = {20},
	urldate = {2026-07-07},
	journal = {Physical Review Letters},
	author = {Iemini, F. and Mazza, L. and Fallani, L. and Zoller, P. and Fazio, R. and Dalmonte, M.},
	month = may,
	year = {2017},
	pages = {200404},
}

@article{lisandrini_majorana_2022,
	title = {Majorana edge modes in a spinful-particle conserving model},
	volume = {106},
	issn = {2469-9950, 2469-9969},
	url = {https://link.aps.org/doi/10.1103/PhysRevB.106.245121},
	doi = {10.1103/PhysRevB.106.245121},
	language = {en},
	number = {24},
	urldate = {2026-07-07},
	journal = {Physical Review B},
	author = {Lisandrini, Franco T. and Kollath, Corinna},
	month = dec,
	year = {2022},
	pages = {245121},
}

@article{alicea_new_2012,
	title = {New directions in the pursuit of {Majorana} fermions in solid state systems},
	volume = {75},
	issn = {0034-4885},
	url = {https://doi.org/10.1088/0034-4885/75/7/076501},
	doi = {10.1088/0034-4885/75/7/076501},
	language = {en},
	number = {7},
	urldate = {2026-07-07},
	journal = {Reports on Progress in Physics},
	publisher = {IOP Publishing},
	author = {Alicea, Jason},
	month = jun,
	year = {2012},
	pages = {076501},
}

@article{lutchyn_majorana_2010,
	title = {Majorana {Fermions} and a {Topological} {Phase} {Transition} in {Semiconductor}-{Superconductor} {Heterostructures}},
	volume = {105},
	copyright = {http://link.aps.org/licenses/aps-default-license},
	issn = {0031-9007, 1079-7114},
	url = {https://link.aps.org/doi/10.1103/PhysRevLett.105.077001},
	doi = {10.1103/PhysRevLett.105.077001},
	language = {en},
	number = {7},
	urldate = {2026-07-07},
	journal = {Physical Review Letters},
	author = {Lutchyn, Roman M. and Sau, Jay D. and Das Sarma, S.},
	month = aug,
	year = {2010},
	pages = {077001},
}

@article{oreg_helical_2010,
	title = {Helical {Liquids} and {Majorana} {Bound} {States} in {Quantum} {Wires}},
	volume = {105},
	copyright = {http://link.aps.org/licenses/aps-default-license},
	issn = {0031-9007, 1079-7114},
	url = {https://link.aps.org/doi/10.1103/PhysRevLett.105.177002},
	doi = {10.1103/PhysRevLett.105.177002},
	language = {en},
	number = {17},
	urldate = {2026-07-07},
	journal = {Physical Review Letters},
	author = {Oreg, Yuval and Refael, Gil and Von Oppen, Felix},
	month = oct,
	year = {2010},
	pages = {177002},
}

@article{jiang_majorana_2011,
	title = {Majorana {Fermions} in {Equilibrium} and in {Driven} {Cold}-{Atom} {Quantum} {Wires}},
	volume = {106},
	copyright = {http://link.aps.org/licenses/aps-default-license},
	issn = {0031-9007, 1079-7114},
	url = {https://link.aps.org/doi/10.1103/PhysRevLett.106.220402},
	doi = {10.1103/PhysRevLett.106.220402},
	language = {en},
	number = {22},
	urldate = {2026-07-07},
	journal = {Physical Review Letters},
	author = {Jiang, Liang and Kitagawa, Takuya and Alicea, Jason and Akhmerov, A. R. and Pekker, David and Refael, Gil and Cirac, J. Ignacio and Demler, Eugene and Lukin, Mikhail D. and Zoller, Peter},
	month = jun,
	year = {2011},
	pages = {220402},
}

@article{iemini_localized_2015,
	title = {Localized {Majorana}-{Like} {Modes} in a {Number}-{Conserving} {Setting}: {An} {Exactly} {Solvable} {Model}},
	volume = {115},
	copyright = {http://link.aps.org/licenses/aps-default-license},
	issn = {0031-9007, 1079-7114},
	shorttitle = {Localized {Majorana}-{Like} {Modes} in a {Number}-{Conserving} {Setting}},
	url = {https://link.aps.org/doi/10.1103/PhysRevLett.115.156402},
	doi = {10.1103/PhysRevLett.115.156402},
	language = {en},
	number = {15},
	urldate = {2026-07-07},
	journal = {Physical Review Letters},
	author = {Iemini, Fernando and Mazza, Leonardo and Rossini, Davide and Fazio, Rosario and Diehl, Sebastian},
	month = oct,
	year = {2015},
	pages = {156402},
}

@article{lang_topological_2015,
	title = {Topological states in a microscopic model of interacting fermions},
	volume = {92},
	copyright = {http://link.aps.org/licenses/aps-default-license},
	issn = {1098-0121, 1550-235X},
	url = {https://link.aps.org/doi/10.1103/PhysRevB.92.041118},
	doi = {10.1103/PhysRevB.92.041118},
	language = {en},
	number = {4},
	urldate = {2026-07-07},
	journal = {Physical Review B},
	author = {Lang, Nicolai and Büchler, Hans Peter},
	month = jul,
	year = {2015},
	pages = {041118},
}

@article{keselman_gapless_2015,
	title = {Gapless symmetry-protected topological phase of fermions in one dimension},
	volume = {91},
	copyright = {http://link.aps.org/licenses/aps-default-license},
	issn = {1098-0121, 1550-235X},
	url = {https://link.aps.org/doi/10.1103/PhysRevB.91.235309},
	doi = {10.1103/PhysRevB.91.235309},
	language = {en},
	number = {23},
	urldate = {2026-07-07},
	journal = {Physical Review B},
	author = {Keselman, Anna and Berg, Erez},
	month = jun,
	year = {2015},
	pages = {235309},
}

@article{baier_extended_2016,
	title = {Extended {Bose}-{Hubbard} models with ultracold magnetic atoms},
	volume = {352},
	url = {https://www.science.org/doi/10.1126/science.aac9812},
	doi = {10.1126/science.aac9812},
	number = {6282},
	urldate = {2026-07-07},
	journal = {Science},
	publisher = {American Association for the Advancement of Science},
	author = {Baier, S. and Mark, M. J. and Petter, D. and Aikawa, K. and Chomaz, L. and Cai, Z. and Baranov, M. and Zoller, P. and Ferlaino, F.},
	month = apr,
	year = {2016},
	pages = {201--205},
}

@article{trefzger_ultracold_2011,
	title = {Ultracold dipolar gases in optical lattices},
	volume = {44},
	issn = {0953-4075},
	url = {https://doi.org/10.1088/0953-4075/44/19/193001},
	doi = {10.1088/0953-4075/44/19/193001},
	language = {en},
	number = {19},
	urldate = {2026-07-07},
	journal = {Journal of Physics B: Atomic, Molecular and Optical Physics},
	author = {Trefzger, C and Menotti, C and Capogrosso-Sansone, B and Lewenstein, M},
	month = sep,
	year = {2011},
	pages = {193001},
}

@article{cooper_stable_2009,
	title = {Stable {Topological} {Superfluid} {Phase} of {Ultracold} {Polar} {Fermionic} {Molecules}},
	volume = {103},
	copyright = {http://link.aps.org/licenses/aps-default-license},
	issn = {0031-9007, 1079-7114},
	url = {https://link.aps.org/doi/10.1103/PhysRevLett.103.155302},
	doi = {10.1103/PhysRevLett.103.155302},
	language = {en},
	number = {15},
	urldate = {2026-07-07},
	journal = {Physical Review Letters},
	author = {Cooper, N. R. and Shlyapnikov, G. V.},
	month = oct,
	year = {2009},
	pages = {155302},
}

@article{karman_microwave_2018,
	title = {Microwave {Shielding} of {Ultracold} {Polar} {Molecules}},
	volume = {121},
	issn = {0031-9007, 1079-7114},
	url = {https://link.aps.org/doi/10.1103/PhysRevLett.121.163401},
	doi = {10.1103/PhysRevLett.121.163401},
	language = {en},
	number = {16},
	urldate = {2026-07-07},
	journal = {Physical Review Letters},
	author = {Karman, Tijs and Hutson, Jeremy M.},
	month = oct,
	year = {2018},
	pages = {163401},
}

@article{bigagli_observation_2024,
	title = {Observation of {Bose}–{Einstein} condensation of dipolar molecules},
	volume = {631},
	copyright = {2024 The Author(s), under exclusive licence to Springer Nature Limited},
	issn = {1476-4687},
	url = {https://www.nature.com/articles/s41586-024-07492-z},
	doi = {10.1038/s41586-024-07492-z},
	language = {en},
	number = {8020},
	urldate = {2026-07-07},
	journal = {Nature},
	publisher = {Nature Publishing Group},
	author = {Bigagli, Niccolò and Yuan, Weijun and Zhang, Siwei and Bulatovic, Boris and Karman, Tijs and Stevenson, Ian and Will, Sebastian},
	month = jul,
	year = {2024},
	pages = {289--293},
}

@article{chomaz_dipolar_2022,
	title = {Dipolar physics: a review of experiments with magnetic quantum gases},
	volume = {86},
	issn = {0034-4885},
	shorttitle = {Dipolar physics},
	url = {https://doi.org/10.1088/1361-6633/aca814},
	doi = {10.1088/1361-6633/aca814},
	language = {en},
	number = {2},
	urldate = {2026-07-07},
	journal = {Reports on Progress in Physics},
	publisher = {IOP Publishing},
	author = {Chomaz, Lauriane and Ferrier-Barbut, Igor and Ferlaino, Francesca and Laburthe-Tolra, Bruno and Lev, Benjamin L and Pfau, Tilman},
	month = dec,
	year = {2022},
	pages = {026401},
}

@article{bellinato_giacomelli_topology_2026,
	title = {Topology meets superconductivity in a one-dimensional t-{J} model of magnetic atoms},
	volume = {17},
	copyright = {2026 The Author(s)},
	issn = {2041-1723},
	url = {https://www.nature.com/articles/s41467-026-71248-8},
	doi = {10.1038/s41467-026-71248-8},
	language = {en},
	number = {1},
	urldate = {2026-07-07},
	journal = {Nature Communications},
	publisher = {Nature Publishing Group},
	author = {Bellinato Giacomelli, Leonardo and Bland, Thomas and Lafforgue, Louis and Ferlaino, Francesca and Mark, Manfred J. and Barbiero, Luca},
	month = apr,
	year = {2026},
	pages = {5328},
}

@article{fazzini_interaction-induced_2019,
	title = {Interaction-{Induced} {Fractionalization} and {Topological} {Superconductivity} in the {Polar} {Molecules} {Anisotropic} t-{J} {Model}},
	volume = {122},
	issn = {0031-9007, 1079-7114},
	url = {https://link.aps.org/doi/10.1103/PhysRevLett.122.106402},
	doi = {10.1103/PhysRevLett.122.106402},
	language = {en},
	number = {10},
	urldate = {2026-07-07},
	journal = {Physical Review Letters},
	author = {Fazzini, Serena and Barbiero, Luca and Montorsi, Arianna},
	month = mar,
	year = {2019},
	pages = {106402},
}

@article{carroll_observation_2025,
	title = {Observation of generalized t-{J} spin dynamics with tunable dipolar interactions},
	volume = {388},
	url = {https://www.science.org/doi/10.1126/science.adq0911},
	doi = {10.1126/science.adq0911},
	number = {6745},
	urldate = {2026-07-07},
	journal = {Science},
	publisher = {American Association for the Advancement of Science},
	author = {Carroll, Annette N. and Hirzler, Henrik and Miller, Calder and Wellnitz, David and Muleady, Sean R. and Lin, Junyu and Zamarski, Krzysztof P. and Wang, Reuben R. W. and Bohn, John L. and Rey, Ana Maria and Ye, Jun},
	month = apr,
	year = {2025},
	pages = {381--386},
}

\appendix
\section{Appendix A: Scaling analysis of the revival}
\begin{figure}[ht]
    \centering
    \captionsetup[subfigure]{labelformat=empty}
    \subfloat{\includegraphics[width=\columnwidth, valign=c]{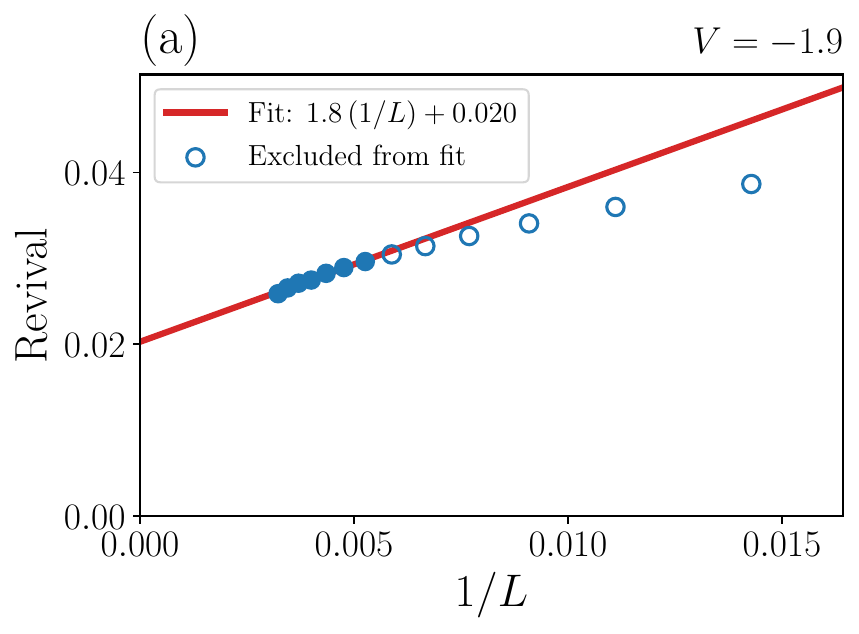}\label{fig:scaling_strong}}
    \vspace{-2em}
    \subfloat{\includegraphics[width=\columnwidth, valign=c]{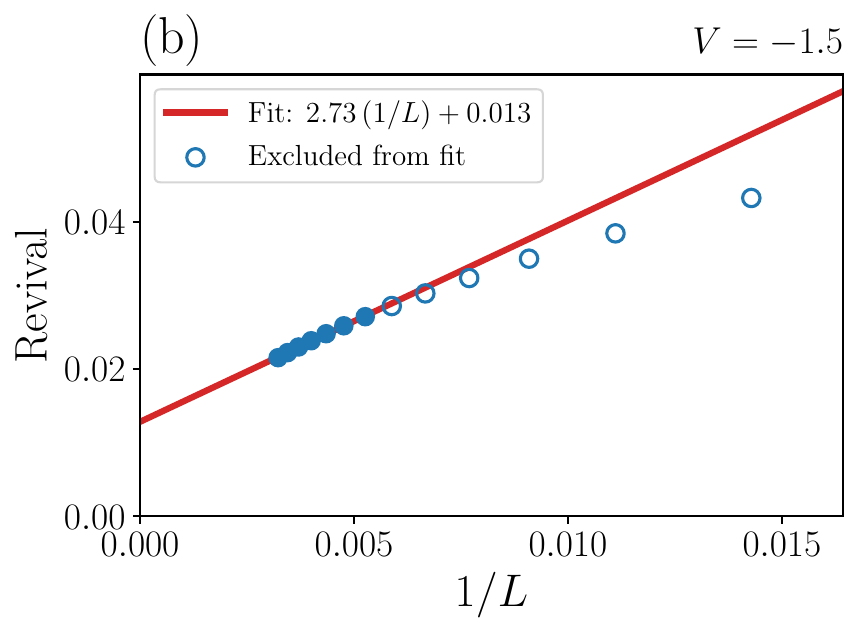}\label{fig:scaling_weak}}

    \caption{\protect\subref{fig:scaling_strong} Revival as a function of the system size $1/L$ at half-filling $N/L=1/2$ for strong attractive interactions $V=-1.9$. The revival is quantified as the maximum value of the two-point correlator $\expvalue{\hat{c}^\dagger_1 \hat{c}_j}$ for $j$ in the range $[L/2, L]$, which corresponds to the right half of the chain. The data is fitted with a function of the form $y = m/L + q$, and from the fit we extract the asymptotic value $q=0.020$ of the revival in the thermodynamic limit $L\to\infty$. \protect\subref{fig:scaling_weak} Same scaling analysis as in Fig.~\ref{fig:scaling_strong}, but for weaker attractive interactions $V=-1.5$. From the fit we extract the asymptotic value $q=0.013$ of the revival in the thermodynamic limit.}
\end{figure}

Since the revival in the $\expvalue{\hat{c}^\dagger_1 \hat{c}_j}$ correlator is a manifestation of the underlying Majorana modes present in the system, it is crucial to show that the revival is not a finite-size effect. To this end, we perform a scaling analysis of the revival as a function of the system size $L$. We focus on the half-filling $N/L=1/2$ case, where $N$ is the total number of fermions. Furthermore, we choose values of $L$ of the form $L=4n+2$ with $n\in\mathbb{N}$, such that the system is always half-filled with an odd number of fermions $N=2n+1$ and the sign of the revival is always positive (see Fig.~\ref{fig:revival_N}). More precisely, the revival is quantified as the maximum value of the two-point correlator $\expvalue{\hat{c}^\dagger_1 \hat{c}_j}$ for $j$ in the range $[L/2, L]$, which corresponds to the right half of the chain.

In Fig.~\ref{fig:scaling_strong}, we plot the revival as a function of the system size $L$ for strong attractive interactions $V=-1.9$ and perform a fit of the form $y =m/L + q$ to the data. From the fit, we extract the asymptotic value $q=0.020$ of the revival in the thermodynamic limit $L\to\infty$ at half-filling, which is non-zero and thus confirms that the revival is not a finite-size effect.

In Fig.~\ref{fig:scaling_weak}, we perform the same scaling analysis for weaker attractive interactions $V=-1.5$. From the fit, we extract the asymptotic value $q=0.013$ of the revival in the thermodynamic limit at half-filling. This confirms that the revival persists in the infinite-size limit for a wide range of interaction strengths $V$ also further away from the phase transition at $V_c=-2$.

\section{Appendix B: Revival at different fillings}
\begin{figure}[ht]
    \centering
    \includegraphics[width=\columnwidth]{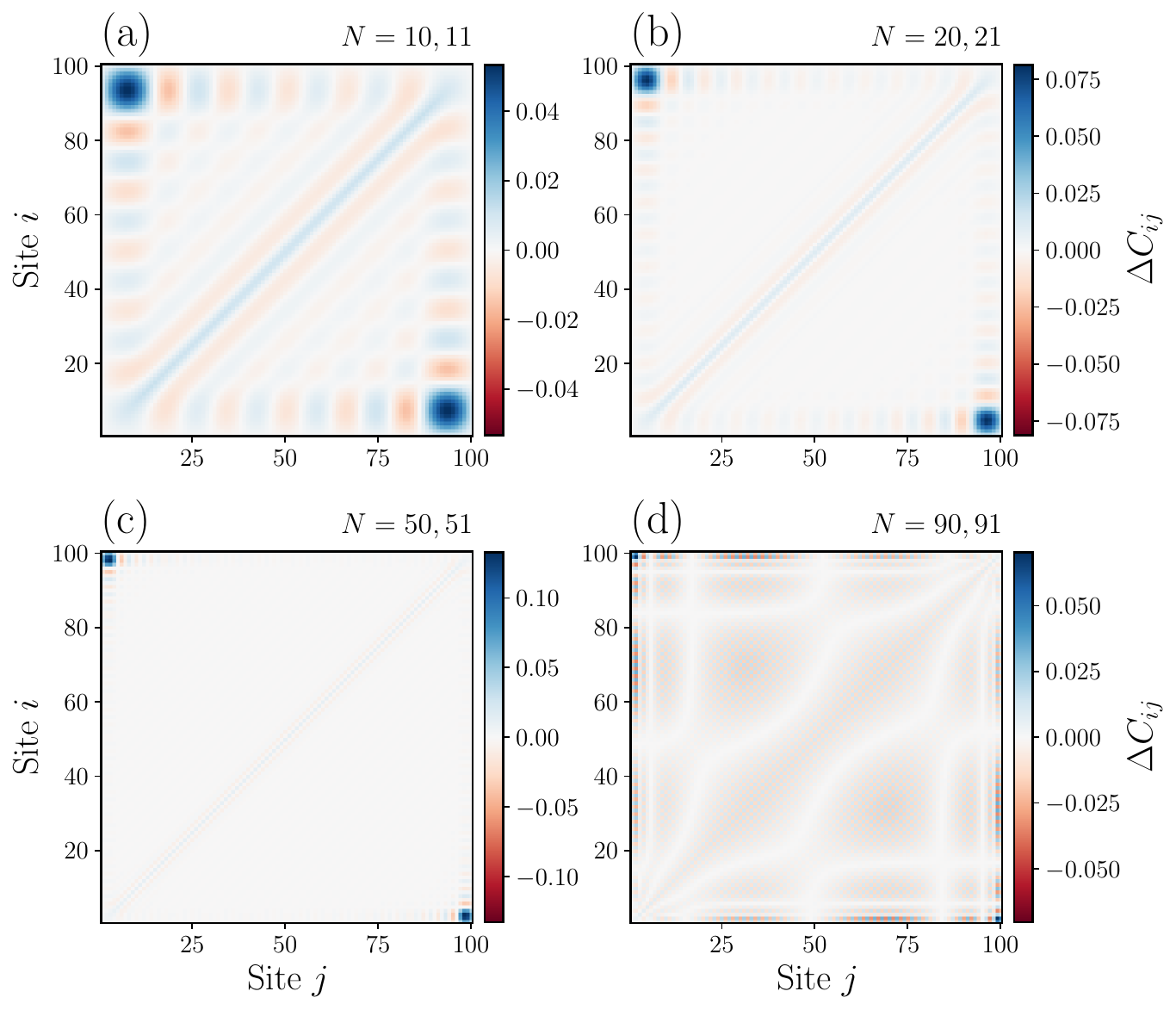}
    \caption{Matrix $\Delta C_{ij} = \bra{N+1} \hat{c}_i^\dagger \hat{c}_j \ket{N+1} - \bra{N} \hat{c}_i^\dagger \hat{c}_j \ket{N}$ for different fillings $N/L$ at strong attractive interactions $V=-1.9$ and $L=100$. The edge-to-edge revival is visible as a blue peak in the bottom-right and top-left corners of the $\Delta C_{ij}$ matrices. The spatial size of the revival shrinks as the filling is increased and its contrast over the bulk oscillations is enhanced at half-filling $N/L=1/2$.}
    \label{fig:fillings}
\end{figure}
In the main text we showed that the model in Eq. \eqref{model} at half-filling $N/L=1/2$ exhibits a revival in the two-point correlator $\expvalue{\hat{c}^\dagger_1 \hat{c}_j}$. In this Appendix we show that the revival is also present at different fillings $N/L$ with a spatial extension which is comparable to the Fermi wavelength of the system. To capture both the magnitude and the spatial extension of the revival, we consider the following matrix
\begin{align}
    \Delta C_{ij} = \bra{N+1} \hat{c}_i^\dagger \hat{c}_j \ket{N+1} - \bra{N} \hat{c}_i^\dagger \hat{c}_j \ket{N},
\end{align}
where $\ket{N}$ and $\ket{N+1}$ are the ground states of the system with $N$ and $N+1$ fermions, respectively \cite{thomas-markarian_majorana_2025}. In Fig.~\ref{fig:fillings}, we display the matrix $\Delta C_{ij}$ for four different fillings $N/L$. Since the two ground states $\ket{N}$ and $\ket{N+1}$ differ by the parity of the number of fermions, the revival has opposite signs in the two ground states (see Fig.~\ref{fig:revival_N}). For this reason, we expect the matrix $\Delta C_{ij}$ to have a peak when $i$ and $j$ are located at the opposite edges of the chain. This peak is visible in blue color in Fig.~\ref{fig:fillings}, at the bottom-right and top-left corners of the $\Delta C_{ij}$ matrices. We notice that the spatial size of the revival shrinks as the filling is increased, and it is comparable to the Fermi wavelength of the system. Furthermore, bulk oscillations in the $\Delta C_{ij}$ are stronger for fillings away from $1/2$. From an experimental perspective, this makes half-filling a promising regime to observe the revival, since the contrast of the revival over the bulk oscillations is enhanced.

\section{Appendix C: Revival in the excited states}
In this Appendix we show that the revival in the two-point correlator $\expvalue{\hat{c}^\dagger_1 \hat{c}_j}$ is also present in the low-lying excited states of the system. This is a result that is both conceptually and experimentally relevant, since it shows that the revival is a robust feature of the system which is not limited to the ground state and that it can be observed in a thermal mixture of low-energy states.

Since for $V>-2$ the model in Eq. \eqref{model} is a gapless Luttinger liquid, the low-energy excited states emerge on top of the ground state as long-wavelength density fluctuations and the two-point correlator $\expvalue{\hat{c}^\dagger_i \hat{c}_j}$ can be studied using bosonization techniques. In the following, we take the continuum limit and consider periodic boundary conditions \cite{giamarchi_quantum_2003}. A posteriori, we will see that these two simplifying assumptions do not affect the main conclusion of this Appendix.

In the bosonization framework, the low-energy properties of the system can be obtained by confining ourselves close to the Fermi points $k_F$ and $-k_F$ so that the fermionic field operator can be decomposed into
\begin{align}
    \hat\psi(x) &\approx \frac{1}{\sqrt{L}} \sum_{k\sim k_F} e^{ikx} \hat{c}_k + \frac{1}{\sqrt{L}}\sum_{k\sim -k_F} e^{ikx} \hat{c}_k\\
    &\equiv \hat\psi_R(x) + \hat\psi_L(x)
\end{align}
where $\hat\psi_R(x)$ and $\hat\psi_L(x)$ are the right- and left-moving components of the fermionic field operator, respectively. In particular, one can also introduce the densities of right- and left-movers $\hat\rho_R(x) = \hat\psi_R^\dagger(x)\hat\psi_R(x)$ and $\hat\rho_L(x) = \hat\psi_L^\dagger(x)\hat\psi_L(x)$. From the Fourier components of the right- and left-moving densities, $\hat\rho_{R}(p)$ and $\hat\rho_{L}(p)$, one can build the following bosonic operators:
\begin{align}
    \hat{b}^\dagger_p = \sqrt{\frac{2\pi}{L|p|}} \hat\rho^\dagger_{R(L)}(p) \quad \text{for } p>0\quad(p<0).
\end{align}
We now introduce the $r$ label, which takes the values $r=1$ and $r=-1$ for right- and left-moving fermions, respectively. It can be shown that the fermionic field operators are displacement operators of the bosonic modes $\hat{b}_p$ and $\hat{b}_p^\dagger$:
\begin{align}
    \hat\psi_r(x) &= \frac{1}{\sqrt{2 \pi \gamma}} e^{i r k_F x} \exp\left( \sum_{p} \beta_p^{(r)}(x) \hat{b}_p^\dagger - \beta_p^{*(r)}(x) \hat{b}_p \right)
\end{align}
where
\begin{align}
    \beta_p^{(r)}(x) = -\sqrt{\frac{2\pi}{L}} \frac{\theta(rp)}{\sqrt{|p|}} e^{-ipx}e^{- \gamma |p|},
\end{align}
$\theta(\cdot)$ is the Heaviside step function and $\gamma$ is an unimportant UV regularization parameter. The low-energy Hamiltonian for Luttinger liquids with Luttinger parameter $K$ and renormalized velocity $u$ is given by
\begin{align}
    \label{eq:luttinger_hamiltonian}
    \hat{H} = \frac{u}{4} \sum_{p} |p| \begin{pmatrix} \hat{b}_p & \hat{b}_{-p}^\dagger \end{pmatrix} \begin{pmatrix} \frac{1}{K}+K & \frac{1}{K}-K \\ \frac{1}{K}-K & \frac{1}{K}+K \end{pmatrix} \begin{pmatrix} \hat{b}_p^\dagger \\ \hat{b}_{-p} \end{pmatrix}.
\end{align}
The Hamiltonian in Eq. \eqref{eq:luttinger_hamiltonian} is quadratic and can be diagonalized by a Bogoliubov rotation
\begin{align}
    \hat{b}_p^\dagger = \cosh(\varphi) \hat{a}_p^\dagger + \sinh(\varphi) \hat{a}_{-p},
\end{align}
where the $\hat{a}_p$ operators are bosonic and the Bogoliubov angle $\varphi$ is related to the Luttinger parameter $K$ by $\tanh(2\varphi) = (K^2-1)/(K^2+1)$. The fermionic field operators rewritten in terms of the Bogoliubov modes $\hat{a}_p$ and $\hat{a}_p^\dagger$ are still displacement operators
\begin{align}
    \hat\psi_r(x) &= \frac{1}{\sqrt{2 \pi \gamma}} e^{i r k_F x} \exp\left( \sum_{p} \alpha_p^{(r)}(x) \hat{a}_p^\dagger - \alpha_p^{*(r)}(x) \hat{a}_p \right)
\end{align}
where
\begin{align}
    \alpha_p^{(r)}(x) = -\sqrt{\frac{2\pi}{L|p|}} e^{-ipx} \left[\theta(rp) \cosh(\varphi) - \theta(-rp) \sinh(\varphi)\right].
\end{align}
From the revival features of the model in Eq. \eqref{model}, we know that the ground state $\ket{N}$ for $N$ fermions is not described exactly by Luttinger liquid theory. Nevertheless, the revival is a short-distance, edge-localized feature and for this reason, we still expect the low-lying excited states to be well described by states of the form
\begin{align}
    \label{eq:excited_states}
    \hat{a}^\dagger_p \ket{N}
\end{align}
for small wavevectors $p$. This corresponds to long-wavelength density fluctuations on top of the ground state.

To understand if the revival persists in the low-lying excited states described by Eq. \eqref{eq:excited_states}, we consider two-point correlators of the form $\ev{\hat{a}_p \hat\psi_R(x) \hat\psi_R^\dagger(0) \hat{a}^\dagger_p}{N}$ and examine how this correlator differs from the ground-state correlator $\ev{\hat\psi_R(x) \hat\psi_R^\dagger(0)}{N}$. The operator appearing in the excited-state correlator is of the form $\hat{a} \hat{D}(\alpha) \hat{D}(\beta) \hat{a}^\dagger$, where $\hat{D}(\alpha)$ is a displacement operator. Using the following identity for single-mode displacement operators
\begin{widetext}
\begin{align}
    \hat{a} \hat{D}(\alpha) \hat{D}(\beta) \hat{a}^\dagger &= \hat{D}(\alpha + \beta)(1-|\alpha+\beta|^2)+ \hat{D}(\alpha + \beta) \hat{a}^\dagger \hat{a} + (\alpha+\beta) \hat{a}^\dagger \hat{D}(\alpha + \beta).
\end{align}
\end{widetext}
and the fact that the occupation number of the Bogoliubov modes $\hat{a}_p$ is essentially zero when evaluated on the ground state $\ev{\hat{a}_p^\dagger \hat{a}_p}{N} \approx 0$, we arrive at the final result
\begin{widetext}
\begin{align}
    \label{eq:excited_states_correlator}
    \ev{\hat{a}_p \hat\psi_R(x) \hat\psi_R^\dagger(0) \hat{a}_p^\dagger}{N} \approx \ev{\hat\psi_R(x) \hat\psi_R^\dagger(0)}{N}
    \begin{cases}
        1- \frac{2\pi}{L|p|} (K + 1/K + 2) \sin^2[px/2] & \text{for } p > 0\\
        1- \frac{2\pi}{L|p|} (K + 1/K - 2) \sin^2[px/2] & \text{for } p < 0
    \end{cases}.
\end{align}
\end{widetext}
This means that for Luttinger liquids, the low-lying excited states are described by long-wavelength density fluctuations on top of the ground state (see Eq. \eqref{eq:excited_states}) and the two-point correlator evaluated on these excited states is given by the ground-state correlator multiplied by a long-wavelength modulation factor (see Eq. \eqref{eq:excited_states_correlator}). For this reason, the revival in the two-point correlator $\expvalue{\hat{c}^\dagger_1 \hat{c}_j}$ is also present in the low-lying excited states of the system, since the long-wavelength modulation given by the density fluctuations does not affect the short-wavelength physics of the revival.
\end{document}